\documentclass[%
 reprint,
 amsmath,amssymb,
 aps,
]{revtex4-2}

\usepackage{graphicx}
\usepackage{dcolumn}
\usepackage{bm}

\usepackage{braket}
\usepackage{tikz}
\usepackage{float}
\usepackage{todonotes}

\newcommand{\myeq}[1]{\begin{eqnarray}\begin{aligned}#1\end{aligned}\end{eqnarray}}

\newcommand{\fa}{{}^\forall}

\begin{document}

\preprint{APS/123-QED}

\title{Testing Scalable Bell Inequalities for Quantum Graph States \\ on IBM Quantum Devices} 
\thanks{Presented at the poster session in QIP Conference 2021.}

\author{Bo Yang$^1$}
\author{Rudy Raymond$^{2,3}$}
\author{Hiroshi Imai$^1$}
\author{Hyungseok Chang$^4$}
\author{Hidefumi Hiraishi$^5$}
\affiliation{%
 $^1$Graduate School of Information Science and Technology, The University of Tokyo, Bunkyo-ku, Tokyo 113-8656, Japan\\
 $^2$IBM Quantum, IBM Research-Tokyo, 19-21 Nihonbashi Hakozaki-cho, Chuo-ku, Tokyo, 103-8510, Japan\\
 $^3$Quantum Computing Center, Keio University, Hiyoshi 3-14-1, Kohoku-ku, Yokohama 223-8522, Japan\\
 $^4$Graduate School of Education, The University of Tokyo, Bunkyo-ku, Tokyo, 113-8654, Japan\\
 $^5$Graduate School of Science and Technology, Nihon University, Chiyoda-ku, Tokyo, 101-8308, Japan
}%


\date{June 6, 2022}

\begin{abstract}
Testing and verifying imperfect multi-qubit quantum devices are important as such noisy quantum devices are widely available today. 
Bell inequalities are known useful for testing and verifying the quality of the quantum devices from their nonlocal quantum states and local measurements. 
There have been many experiments demonstrating the violations of Bell inequalities but they are limited in the number of qubits and the types of quantum states.  
We report violations of Bell inequalities on IBM Quantum devices based on the scalable and robust inequalities maximally violated by graph states as proposed by Baccari et al. (Ref.\cite{PhysRevLett.124.020402}).
The violations are obtained from the quantum states of path graphs up to 57 and 21 qubits on a 65-qubit and two 27-qubit IBM Quantum devices respectively, and from those of star graphs up to 11 qubits with quantum readout error mitigation (QREM).
We are able to show violations of the inequalities on various graph states by constructing low-depth quantum circuits producing them, and by applying the QREM technique.
We also point out that quantum circuits for star graph states of size $N$ can be realized with circuits of depth $O(\sqrt N)$ on subdivided honeycomb lattices which are the topology of the 65-qubit IBM Quantum device. 
Our experiments show encouraging results on the ability of existing quantum devices to prepare entangled quantum states, and provide experimental evidences on the benefit of scalable Bell inequalities for testing them.
\end{abstract}

\maketitle


\section{\label{sec:level1}Introduction\protect\lowercase{}}

\par Nonlocality of quantum states--first discovered by John S. Bell~\cite{PhysicsPhysiqueFizika.1.195}--is an intriguing consequence of quantum mechanics in which correlations among quantum bits cannot be explain by classical statistics. 
In particular, the nonlocality implies the so-called Bell inequalities that are violated by entangled (or, nonlocal) quantum states but not by any classical (or, local) correlation. 
There is a variety of concepts and experimental tools developed for demonstrating the violation of Bell inequalities~\cite{RevModPhys.86.419}. 
One of them is the CHSH inequality~\cite{PhysRevLett.23.880}, that can be used to test the nonlocality of two qubits. 
There have been many researches extending the CHSH inequality, such as that by Ito, Imai, and Avis~\cite{PhysRevA.73.042109} whose inequality allows wider range of quantum states to violate the classical bounds, or the CHSH-like Bell inequalities for more-than-two quantum bits, such as, the Mermin's inequality for Greenberger–Horne–Zeilinger (GHZ) state~\cite{PhysRevLett.65.1838} and the Bell inequality for graph states~\cite{G_hne_2005}.

\par The Bell inequalities soon find their applications for witnessing entanglement~\cite{Tura_2014, Tura_2015} and for self-testing~\cite{Mayers_Yao_2004, PhysRevLett.97.120405} quantum devices. 
The latter is useful for certifying the \textit{quantumness} of the devices by statistical tests on the correlations resulting from the quantum states they produce without the knowledge of their internal functions. 
However, most of the existing Bell inequalities require measuring correlations on quantum graph states whose number scales exponentially~\cite{Schmied_2016, G_hne_2005} or polynomially~\cite{Tura_2014poly, PhysRevX.7.021042, PhysRevX.7.021005} with the number of qubits involved. 
Recently, Baccari et al.~\cite{PhysRevLett.124.020402} proposed a family of CHSH-like Bell inequalities that are both scalable and robust. The scalability comes from the fact that the new inequalities can be tested by measuring correlations 
on quantum graph states whose number scales only linearly with the number of qubits. 
The robustness stems from the fact that the maximal violation is obtained from quantum graph states whose ratio of the quantum bound against the classical bound tends to a constant for sufficiently large number of qubits. 
In addition, the fidelity of the violating quantum states against the corresponding quantum graph states is a linear function of the magnitude of the violations. 
The scalability and robustness of the new CHSH-like Bell inequalities are therefore potential for self-testing noisy quantum devices available today. 

\par We have been witnessing the proliferation of near-term quantum devices~\cite{jurcevic2020demonstration, pino2020demonstration, arute2020quantum, arute2020, Qiskit}.
As in January 2021 there are at least seventeen multi-qubit quantum devices made available at IBM Quantum Experience~\footnote{IBM Quantum Experience, https://quantum-computing.ibm.com/}. 
Although far from perfect, their number and quality of the qubits has been much improved since the first introduction of their predecessor in 2016. 
The quantum devices with the largest number of qubits is the 65-qubit (\verb+ibmq_brooklyn+) followed by other smaller devices. 
The quality of those devices is measured with the \textit{Quantum Volume}~\cite{Cross_2019,jurcevic2020demonstration} which is a single metric incorporating the number of qubits and the depth of the quantum circuits applicable to the qubits before they decohere. 
Those devices offer testbeds for investigating the quantum states they produce, i.e., to see if such topologically-limited noisy devices can entangle more qubits and in which way. 
For example, Wei et al.~\cite{Wei2019} experimentally demonstrated the ability to produce GHZ states up to 18 qubits on a 20-qubit IBM Quantum device measured by their proposed scalable entanglement metric. 
González et al.~\cite{Gonz_lez_2020} and Huang et al.~\cite{Huang_2020} used Mermin-type Bell inequalities to confirm entanglement of GHZ states up to 5 qubits. 
Nevertheless, the previous experiments are limited and are also difficult to verify different types of entangled quantum states that may depend on the underlying quantum devices. 

\par We address the task of testing noisy quantum devices with various quantum graph states based on the the family of CHSH-like Bell inequalities of Baccari et al.~\cite{PhysRevLett.124.020402}.
We exploit their Bell inequalities to construct various inequalities based on the qubit layout topology of the underlying quantum devices. The inequalities are maximally violated by quantum graph states, whose graphs can be varied based on the connectivity of the qubits of the devices.
We construct path graphs, star graphs, and the qubit-connection graph corresponding to the physical qubit connectivity
on the QV32, 65-qubit 
\verb+ibmq_brooklyn+ device,
the QV32, 27-qubit
\verb+ibmq_cairo+,
\verb+ibmq_hanoi+,
\verb+ibmq_kawasaki+,
\verb+ibmq_mumbai+,
\verb+ibmq_toronto+ devices,
and the QV128, 27-qubit \verb+ibmq_montreal+ device.
Most of these devices appeared in the period from 2020 to 2021.
In particular, we observe the violation of full 65-qubit connection graph states, showing the ability of IBM Quantum devices to create the well entangled large quantum states.
We also observe the maximum violation of path graphs up to 57-qubits and the maximum violation of star graphs up to 11-qubit with the efficient quantum readout error mitigation (QREM) by Yang et al. \cite{yang2022efficient}.
These experimental results also support the applicability of the scalable CHSH-like inequality by Baccar et al. to the entanglement witness and device benchmarking of the growing near-term quantum devices.


\par The violations are made possible by shallow-depth circuits to produce the corresponding graph states. 
Namely, path graphs are from depth 2 quantum circuits as in \cite{ghz_fidelity20mooney}, and star graphs on 5-qubit or larger are from quantum circuits avoiding SWAP gates following a similar construction shown in~\cite{Wei2019}.
We also provide a generalization of constructing quantum circuits of $n$-qubit star graph states with depth $O(\sqrt n)$ on the subdivided honeycomb lattice structure, which is the typical topology of IBM Quantum devices. 

\par The rest of the paper is organized as follows. Section~\ref{sec:settings_of_experiments} explains the experimental settings by introducing the graph states, the corresponding CHSH-like Bell inequalities and the quantum circuits producing the graph states. Section~\ref{sec:results_of_experiments} shows the experimental results on IBM Quantum devices showing their ability to entangle more qubits than reported before. Section~\ref{sec:conclusion} concludes with the discussion of the results and future works.

\section{\label{sec:settings_of_experiments}Settings of Experiments\protect\lowercase{}}

\par In this section, we describe the settings and procedures of our experiments which were implemented on IBM Quantum Experience.
The device information (calibration data) of each quantum device we used is listed in the Appendix \ref{sec:appendix_device_info}.

\subsection{Preliminaries of Graph State \protect\lowercase{}}

\par First, we consider a graph $G=(V, E)$ is a simple undirected graph with vertex set $V=\{1,2,\cdots, N\}$ and edge set $E = \{\{u, v\}| u, v \in V, u \neq v\}$.
Let $n(v)$ be the vertex set of neighbourhoods of the vertex $v$ and $n[v] := n(v)\cup \{v\}$ be the vertex set containing neighbourhoods of the vertex $v$ and $v$ itself.
Given a graph $G=(V, E)$, the graph state $\ket{\psi_G}$ associated to the graph $G$ is defined in the following way.
To every vertex $v$, $G_v$ is an operator on $N$-qubit system written as
\myeq{
G_{v}=\sigma_{X}^{(v)} \otimes \bigotimes_{i \in n(v)} \sigma_{Z}^{(i)},
\label{eq:stabilizer}
}
where the Pauli operator $\sigma_X^{(j)}$ or $\sigma_Z^{(j)}$ acts on the qubit $j$.
Then the graph state $\ket{\psi_G}$ associated to $G$ is defined to be the unique simultaneous eigenvector
\myeq{
\left|\psi_{G}\right\rangle:=\prod_{(i, j) \in E} CZ(i, j)|+\rangle^{\otimes N}.
\label{eq:graph_state}
}
We can prepare the quantum circuit of graph state $\ket {\psi_G}$ according to (\ref{eq:graph_state}).

\subsection{The Scalable Bell Inequality of Baccari et al. \protect\lowercase{}}

\par In order to review the scalable Bell inequality of Baccari et al. used in our experiment, we refer to their original notations in \cite{PhysRevLett.124.020402}.
The inequality we used is the modified version of their original inequality, which is also mentioned in \cite{PhysRevLett.124.020402}.
Using the notation of stabilizer measurement (\ref{eq:stabilizer}), the general form of their modified inequality becomes (\ref{eq:modified_bell_ineq}).
\begin{eqnarray}
\begin{aligned}
I_{G}(F)
=&\sum_{v \in F}\left(\operatorname{deg}(v)\left\langle G_v\right\rangle + \sum_{i \in n(v)}\left\langle G_i\right\rangle\right)\\
+&\sum_{i \notin \bigcup_{v \in F} n[v]}\left\langle G_i\right\rangle \leq \beta_{G}^{C}(F)
\end{aligned}
\label{eq:modified_bell_ineq}
\end{eqnarray}
where $\beta_G^C(F)$ represents the classical bound of graph $G$ to the choice of $F$.
In addition, $F$ satisfies $\fa u, v \in F, \  n[u] \cap n[v]=\emptyset$.
We also define $\beta_G^Q(F)$ as the quantum bound of $G$.

We computed $I_G(F)$ of path graph states with the optimal choice of $F$, obtaining higher ratio of $\beta_G^Q / \beta_G^C$ in order to make the gap between $\beta_G^Q$ and $\beta_G^C$ clearer.
For star graphs, $I_G(F)$ becomes quite simpler and it is described as the following form (\ref{eq:star_bell_ineq}).
{\small
\begin{eqnarray}
I_{S_N}=\sqrt{2} \left((N-1)\left\langle\sigma_{X}^{(1)}\sigma_{Z}^{(2)}\cdots\sigma_{Z}^{(N)}\right\rangle +\sum_{i \in V\backslash \{1\}}\left\langle\sigma_{Z}^{(1)}\sigma_{X}^{(i)}\right\rangle\right)\nonumber \\
\label{eq:star_bell_ineq}
\end{eqnarray}
}

\subsection{Circuit Preparation\protect\lowercase{}}

\par In this experiment, we used IBM Quantum 65-qubit device (\verb+ibmq_brooklyn+) and 27-qubit devices (\verb+ibm_cairo+, \verb+ibm_hanoi+, \verb+ibm_kawasaki+, \verb+ibmq_montreal+, \verb+ibm_mumbai+ and \verb+ibmq_toronto+).
On these devices, we investigated the correlations of path graph $P_N$, star graph $S_N$, and the connection graphs of each device.
In order to prepare shallower circuits, we referred to the circuit designing techniques used by Wei et al. \cite{Wei2019} and Mooney et al. \cite{ghz_fidelity20mooney}.

\subsubsection{Preparing Path Graph State}

\par Path graph state $\ket{\psi_{P_N}}$ can be prepared by shallow circuit with constant depth 2, as shown in \cite{ghz_fidelity20mooney}.
Once we have prepared $\ket+^{\otimes N}$ state, we apply control-Z gate to every other edge of the path.
Then we apply control-Z gate to every other remaining edge on which the control-Z gate was not applied at the previous step.
We used the qubit layout in Fig.~\ref{fig:initial_layout_path} in Appendix \ref{sec:appendix_initial_layout} in order to make as long paths as possible on each device.
We tested path graph states from the size 2 up to the maximum size that can be taken on each device.

\subsubsection{Preparing Star Graph State}

\par Since star graphs are equivalent to GHZ states in terms of local Clifford operations \cite{PhysRevA.69.022316}, star graphs can be made from GHZ state by applying local Hadamard gate to every qubit except for the qubit representing the central node in the graph.
That is, assuming that the central vertex is labled by $1$, the following equation holds.
\myeq{
\ket{\psi_{S_N}} = \left(I\otimes H^{\otimes (N-1)}\right)\ket{\psi_{GHZ_N}}
}


\par Then the remaining task is to prepare GHZ state in a shallower manner, which we can use the technique as shown in \cite{Wei2019}.
The main idea of this technique is that GHZ states can be prepared without qubit swapping operations on any tree structured physical connection of qubits.
GHZ states is realized by applying the Hadamard gate to an initial qubit and then applying the X gate to other qubits controlled by the initial qubit.
Since the qubits of entangled part of GHZ state are all equivalent, we can apply the control-X gates to different pairs of qubits in parallel by properly changing the control qubits.
By doing so, it is possible to realize shallower circuit with depth $O(\sqrt{N})$ for star graph state with size $N$ on the topology of IBM Quantum 65-qubit devices.
In-depth discussion on the proof of this is in the Appendix \ref{sec:appendix_depth_proof}.

\par In our experiments, we prepared the star graph $\ket{\psi_{S_N}}$ of size $N = 2, \cdots, 39$ on 65-qubit devices, and of size $N = 2,\cdots, 27$ on 27-qubit devices.
The details on how we prepared star graphs on each device is shown in Fig.~\ref{fig:initial_layout_star} in Appendix \ref{sec:appendix_initial_layout}.

\par Besides, the grouping of observables with separable measurements, which has been conventionally used, would allow us to run fewer circuits and save resources of quantum computers. 
The idea of this technique is to first measure the sum of commutative observables at once, then extract the expectation value of each observable.
The commutativity of observables in Baccari et al.'s inequality is decided by the stabilizers of the graph state.
The adjacent stabilizers $\sigma_X^{(i)}\prod_{k\in n(i)}$ and $\sigma_X^{(i)}$ for $(i,j)\in E$ are noncommutative due to the noncommutativity of $\sigma_X^{(i)}\sigma_Z^{(j)}$ and $\sigma_Z^{(i)}\sigma_X^{(j)}$.
This property characterizes the minimum number of grouped measurements with the chromatic number of the associated graph $G=(V, E)$ for a graph state $\ket{\psi_G}$.
Since path graphs, star graphs, and connection graphs of the heavy-hexagonal structure are all 2-colorable, we can merge the stabilizer measurement observables into two measurements.
For path graphs, the measurements can be grouped into $\left\langle\sigma_{X}^{(0)}\sigma_{Z}^{(1)}\sigma_{X}^{(2)}\sigma_{Z}^{(3)}\cdots\right\rangle$ and $\left\langle\sigma_{Z}^{(0)}\sigma_{X}^{(1)}\sigma_{Z}^{(2)}\sigma_{X}^{(3)}\cdots\right\rangle$.
For star graphs, the measurements can be grouped into $\left\langle\sigma_{X}^{(0)}\sigma_{Z}^{(1)}\sigma_{Z}^{(2)}\sigma_{Z}^{(3)}\cdots\right\rangle$ and $\left\langle\sigma_{Z}^{(0)}\sigma_{X}^{(1)}\sigma_{X}^{(2)}\sigma_{X}^{(3)}\cdots\right\rangle$.

\subsubsection{Preparing the Graph Structure of Each Device}

\par The graph structure of qubit connection of each quantum devices we used can be seen as a subdivision of honeycomb graph.
Let us define this graph of size $N$ as $TH_N$.
Since the maximum degree of $TH_N$ is 3, it is shown to be 3-edge colorable by Vizing's theorem \cite{vizing1964estimate}.
Therefore, we can prepare the quantum circuit for $TH_N$ in circuit depth 3.
The specific construction of $TH_N$ corresponding to graph structure of each device is explained on Fig.\ref{fig:initial_layout_connection} of Appendix \ref{sec:appendix_initial_layout}.

\begin{figure*}[t]
\includegraphics[width=\linewidth]{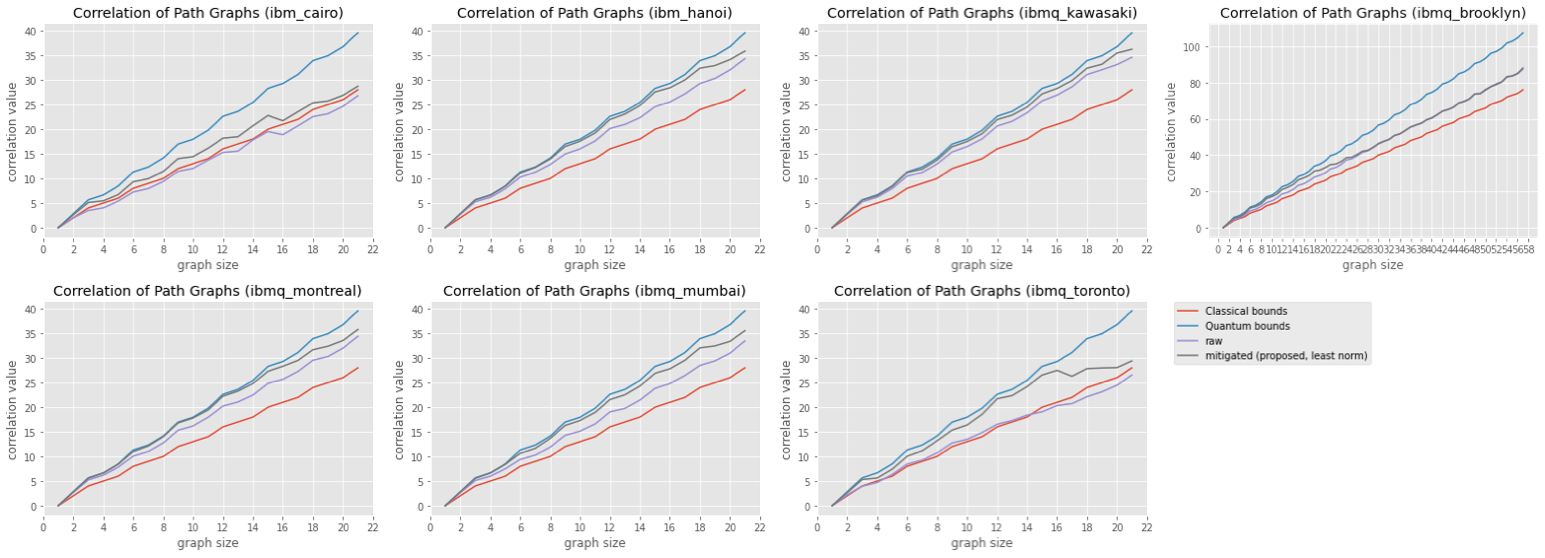}
\caption{\label{fig:bell_path} The correlations of path graphs on each quantum devices. The red lines denote the upper bounds of classical correlation of Baccari et al.'s inequality, and the blue lines denote the upper bounds of quantum correlation of the inequality. The purple and black lines represent the measured correlation on real devices with and without QREM, respectively.}
\end{figure*}
\begin{figure*}[t]
\includegraphics[width=\linewidth]{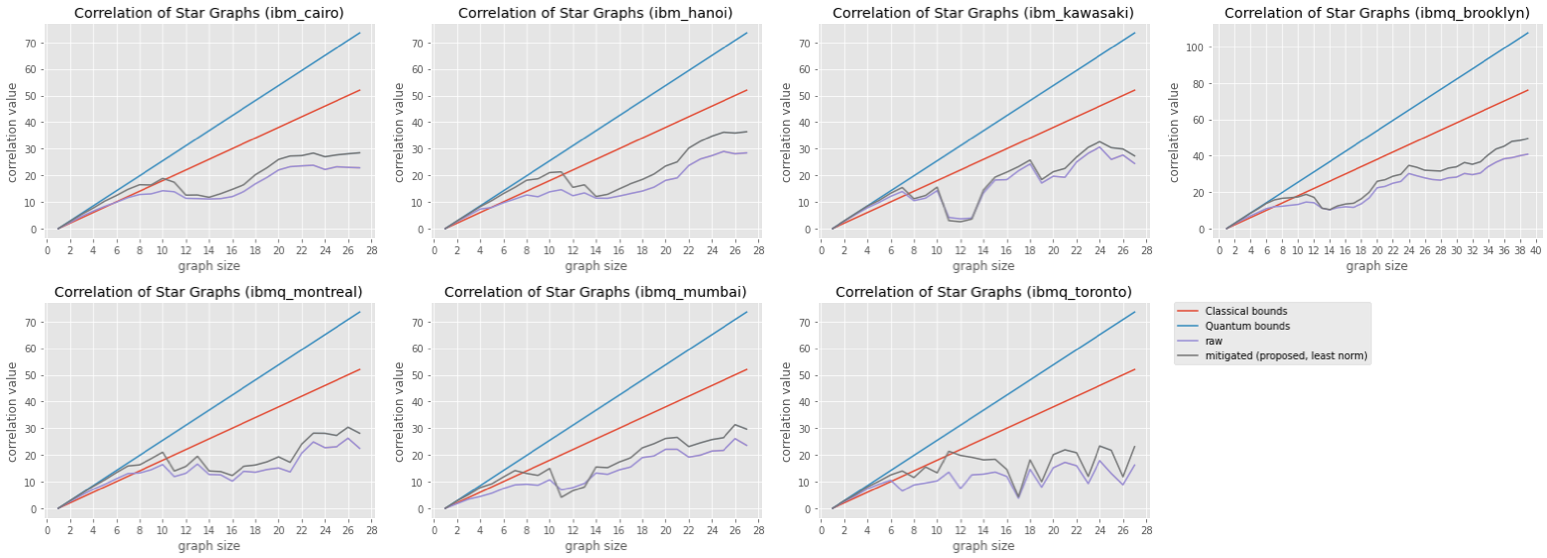}
\caption{\label{fig:bell_star} The correlations of star graphs on each quantum devices. The red lines denote the upper bounds of classical correlation of Baccari et al.'s inequality, and the blue lines denote the upper bounds of quantum correlation of the inequality. The purple and black lines represent the measured correlation on real devices with and without QREM, respectively.}
\end{figure*}
\begin{figure*}[t]
\includegraphics[width=\linewidth]{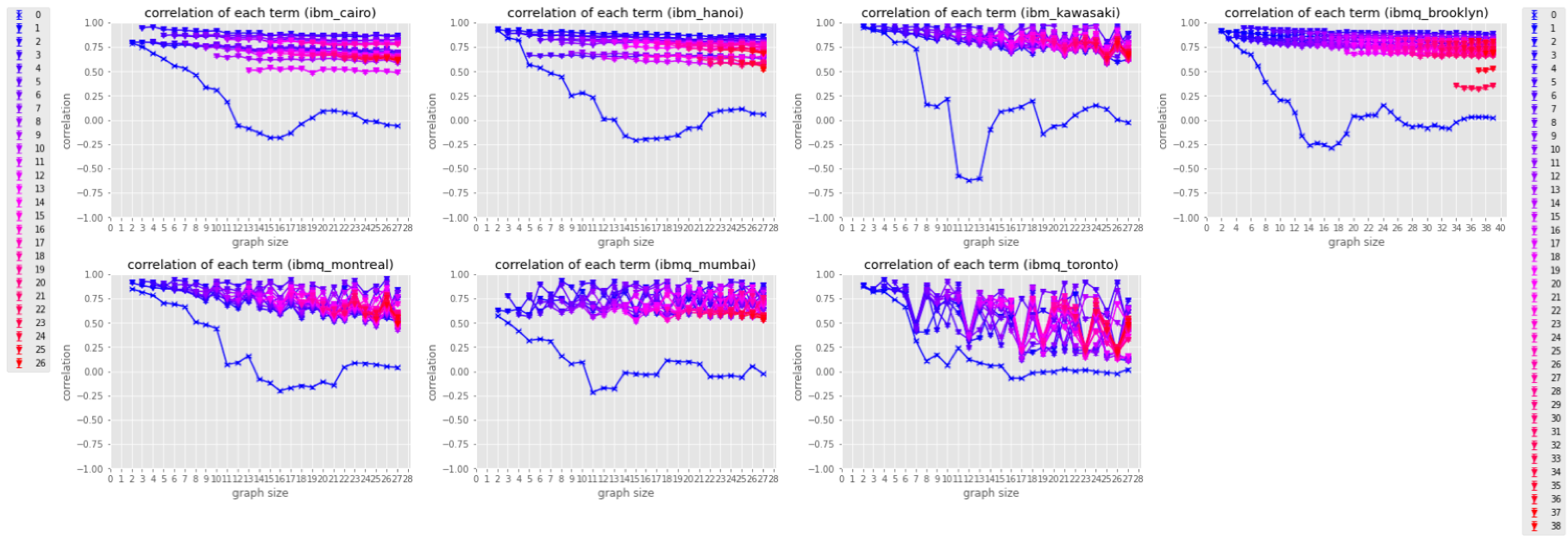}
\caption{\label{fig:bell_star_term_wise} The term-wise correlations of star graphs on each quantum device without QREM.
The labels in the left edge represent the term-wise correlation labels 27-qubit devices.
The labels in the right edge represent the term-wise correlation labels 65-qubit devices (ibmq\_brooklyn only).}
\end{figure*}

\section{\label{sec:results_of_experiments}Results of Experiments\protect\lowercase{}}

\par The python codes of this experiment are stored at \cite{master_thesis}.
The result of each experiment is averaged over 8192 shots. 
By Hoeffding's inequality, the number of shots for each circuits scales in $O\left(\frac {n^2} {\delta^2}\right)$, where $\delta$ is the error tolerance.
In this sense, 8192 shots per circuit are enough for the size of qubits in our experiments.
The quantum correlations for path graphs on each device are shown in Fig. \ref{fig:bell_path}.
The quantum correlations for star graphs on each device are shown in Fig. \ref{fig:bell_star}.
The purple plots are the raw correlations without QREM, and the black lines are the correlations with the proposed QREM method by least norm problem.

\par Fig. \ref{fig:bell_star_term_wise} shows the term-wise mitigated correlations in (\ref{eq:star_bell_ineq}) for each size of star graphs.
For example, the curve labeled by 0 indicates the correlation of the term $\left\langle\sigma_X^{(0)}\sigma_Z^{(1)}\cdots\sigma_Z^{(N-1)}\right\rangle$ in (\ref{eq:star_bell_ineq}) for graph size $N = 2, 3, 4, \cdots$, and the curve labeled by 3 indicates the correlation of the term $\left\langle\sigma_X^{(0)}\sigma_Z^{(3)}\right\rangle$ in (\ref{eq:star_bell_ineq}) for graph size $N = 4, 5, 6, \cdots$.

\par From Fig. \ref{fig:bell_path}, we observe the path graphs violate the inequality with a clear gap from classical bounds without QREM on most of the devices, while the plots of \verb+ibm_cairo+ and \verb+ibmq_toronto+ show the worse correlations that do not violate the classical upper bounds.
However, when the QREM method is added, all plots of correlations on each device violated the classical bounds.
With QREM, the measured correlations on the devices except for \verb+ibm_cairo+ almost reach the theoretical upper bound of quantum correlation.
Besides, thanks to the constant-depth circuit preparation of path graph states, the plots of path graphs on each device seem to grow stably to the system size between the classical bound and quantum bound.
Therefore, we may say path graphs can be well prepared on the current IBM Quantum devices.

\par As for the star graph states, the raw correlations without QREM only violate the classical upper bounds up to 4 to 6 qubits on each device.
With the proposed QREM, the maximum size of violation increased to size 11 on \verb+ibm_hanoi+ and \verb+ibmq_toronto+, to size 10 on \verb+ibm_cairo+ and \verb+ibmq_montreal+, to size 9 on \verb+ibmq_brooklyn+, and to size 7 on \verb+ibm_kawasaki+ and \verb+ibm_mumbai+.
The decreases of the total correlations and the term-wise correlations of the central vertex (qubit 0) of graph state may reflect the increase of circuit depth, which would make each qubit more vulnerable to decoherence.
This is obvious especially between the size 10 and size 11.

\par The figures of term-wise correlation in Fig. \ref{fig:bell_star_term_wise} show more intriguing properties of correlations.
Although the observable for the central qubit and those for the other qubits are computed by two grouped measurements of the same number of qubits, only the correlations of the first term get worse as the graph size increases, even taking the negative values.
One direct reason is that the grouped measurement of the other qubits other than the central qubit changes to the star graph state into GHZ state.
Since GHZ state has dominant probability in a few states of $\ket{00\ldots 0}$ and $\ket{11\ldots 1}$ by the measurement with computational bases, the outcome probability distribution may be less affected by decoherence and other noises.


\begin{table}[htb]
\centering
\caption{\label{tab:bell_connection_graph} 
The correlation values of the whole-qubit honeycomb graph on each device.
The $\beta^C_G$ and $\beta^Q_G$ represents the theoretical upper bounds of classical correlation and quantum correlation respectively.
The "raw $\overline{\beta}^Q_G$" and "mitigated $\overline{\beta}^Q_G$" denotes the measured quantum correlations without and with QREM respectively.
Here the proposed the QREM with least norm method is used.
}
\begin{tabular}{ccccc}
\hline
device & $\beta^C_G$ & $\beta^Q_G$ & raw $\overline{\beta}^Q_G$ & mitigated $\overline{\beta}^Q_G$\\ \hline
\verb+ibmq_brooklyn+ & $88$ & $121.97$ & $94.58 \pm 0.12$ & $96.79 \pm 0.120$ \\
\verb+ibm_cairo+ & $36$ & $50.08$ & $41.05 \pm 0.075$ & $41.39 \pm 0.07$ \\
\verb+ibm_hanoi+ & $36$ & $50.08$ & $44.21 \pm 0.06$ & $44.61 \pm 0.06$ \\
\verb+ibm_kawasaki+ & $36$ & $50.08$ & $44.74 \pm 0.06$ & $45.39 \pm 0.05$ \\
\verb+ibmq_montreal+ & $36$ & $50.08$ & $38.54 \pm 0.08$ & $38.82 \pm 0.08$ \\
\verb+ibm_mumbai+ & $36$ & $50.08$ & $39.94 \pm 0.08$ & $40.86 \pm 0.08$ \\
\verb+ibmq_toronto+ & $36$ & $50.08$ & $30.52 \pm 0.10$ & $32.21 \pm 0.10$ \\
\hline
\end{tabular}
\end{table}

\par We also report the violation of the CHSH-like Bell inequality by Baccari et al. in the subdivided honeycomb graph using whole qubits on both the 65-qubit device and the 27-qubit devices.
The correlations of full connection graph state associated with the qubit structure of each device are listed in Table \ref{tab:bell_connection_graph}.
This result implies that these IBM Quantum devices except for \verb+ibmq_toronto+ have the ability to prepare a large graph state unique to its qubit layout even using its whole qubits, in rather good accuracy.
Especially the 65-qubit connection graph states on \verb+ibmq_brooklyn+ achieves the largest graph state preparation on IBM Quantum Experience.


\section{\label{sec:conclusion}Conclusion\protect\lowercase{}}

\par Through our experiments, we support the benefits of the CHSH-like inequality proposed by Baccari et al. \cite{PhysRevLett.124.020402} in terms of its scalability and robustness.
The linear-scale increase of measurement terms to the graph size enables us to compute correlation of large graph states on IBM Quantum devices such as \verb+ibmq_brooklyn+ with 65-qubits. 
Bell inequalities that require measuring only constant number of correlations~\cite{Tura_2014,Tura_2015} have been used for experimenting with much larger systems~\cite{Schmied_2016,PhysRevLett.118.140401}.  

\par Using Baccari et al.'s remarkable Bell inequality, we also support the ability of existing IBM Quantum devices to prepare well-entangled large graph states on them.
We report in this work the violation of the inequality for several graph states with a large number of qubits.
Using shallow circuits with depth 2 \cite{ghz_fidelity20mooney}, we have seen path graphs violated the inequality up to the maximum size on each IBM Quantum device.
In particular, for the IBM Quantum 65-qubit device, path graphs showed its quantumness up to size 57.
We also checked the violation of classical bounds for the graph state corresponding to the graph structure of each quantum devices with its whole qubits.
Although the maximum size of star graphs violating the inequality (\ref{eq:modified_bell_ineq}) is rather small compared to the violations in path graphs, our result reports the violation of star graphs up to size 6.
Our preliminary efforts applying QREM \cite{yang2022efficient} showed that the size could be increased maximally up to 11.

\par For future works, one of the possible improvements of circuit preparation can be found in the experiments by Wei et al. \cite{Wei2019}.
During their experiments, they added a collective $\pi$-pulse on all qubits in order to refocus low frequency noise and reduces dephasing errors using the idea of Hahn echo \cite{PhysRev.80.580}.
As they applied the $\pi$-pulse between the entangle process and disentangle process of GHZ states which undo the entangle process, $\pi$-pulse becomes most effective for certain time intervals decided by T1/T2 relaxation times.
Since our experiments do not have the structure of symmetry in terms of entangle process and disentangle process, partial insertion of $\pi$-pulse into the entangled qubits might improve the correlations instead of the direct insertion of $\pi$-pulse into the middle of our circuits.
Other ideas of decreasing the dephasing errors, such as dynamic decoupling methods discussed in \cite{Pokharel_2018}, might also help us improve the total correlations of the inequality.

\par In conclusion, our results for the large quantum states greatly owe to the scalability of the Bell inequality proposed by Baccari et al. \cite{PhysRevLett.124.020402} and we experimentally support the usefulness of their inequality as a powerful tool for the entanglement verification of large quantum states and for the benchmarking of upcoming near-term quantum devices.

\begin{acknowledgments}
The results presented in this paper were obtained in part using an IBM Quantum computing system as part of the IBM Quantum Hub at University of Tokyo. 
\end{acknowledgments}

\bibliography{main}

\onecolumngrid

\appendix
\label{sec:appendix}

\section{Creating Star Graphs with Depth $O(\sqrt N)$ \label{sec:appendix_depth_proof}}

\par At the previous part, we have seen that quantum circuit for star graph state $\ket{\psi_{S_N}}$ is prepared via the GHZ state which can avoid swap operations.
Here we explain why the quantum circuit can be prepared with depth $O(\sqrt N)$ for star graph $S_N$ on the physical qubit layout of \verb+ibmq_brooklyn+.

\par We first describe the construction of tree graph state $\ket{\psi_{T_N}}$ with depth $d$ and see what the physical qubit topology should be taken.
We then show such a graph can be embedded into the topology of the subdivided honeycomb structure.
In order to create a quantum circuit, we start from vertex $1$.
If vertex $1$ is connected with other vertex, say vertex $2$, we can add it to the tree, making $\ket{\psi_{T_2}}$ with depth $1$.

\par Next, if one of the vertices $1,2$ has degree 3 or larger, connected with vertex $3$, and the other vertex has degree 2 or larger, connected with vertex $4$, then we can simultaneously add vertices $3,4$ to vertex $1,2$.
This time, the created tree $\ket{\psi_{T_4}}$ has the depth $2$, with 3 outer vertices on the qubit topology connected to different vertices of $\ket{\psi_{T_4}}$.
Going one step further, if two of three neighborhoods of $\ket{\psi_{T_4}}$ have degree 2 or larger, and the remaining one neighborhood has degree 3 or larger, then we can make $\ket{\psi_{T_7}}$ in one step, and assure 4 additional neighborhoods for $\ket{\psi_{T_7}}$.

\par In this way, the size of tree graph state we can prepare in depth $d$ is $N = \frac 1 2 d(d+1) + 1$.
The condition that the physical qubit topology should satisfy is that they can add $d - 1$ vertices with degree 2, and at least 1 vertex with degree $3$.
Such structure can be found in subdivided honeycomb because every vertex with degree 2 in the subdivided honeycomb is adjacent to vertices with degree 3, and vice versa.
Note that the argument above can be applied to other two-dimensional lattice structures.


\section{Qubit Layout of Quantum Circuits on IBM Quantum Experience \label{sec:appendix_initial_layout}}

\begin{figure}[htb]
  \begin{minipage}[b]{0.29\linewidth}
    \centering
    \includegraphics[keepaspectratio, width=\linewidth]{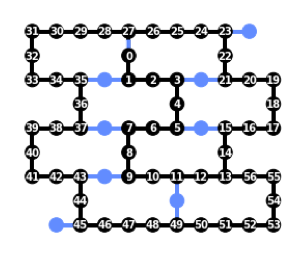}
  \end{minipage}
  \begin{minipage}[b]{0.29\linewidth}
    \centering
    \includegraphics[keepaspectratio, width=\linewidth]{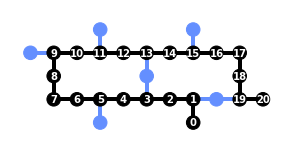}
  \end{minipage}
  \caption{ \label{fig:initial_layout_path}
    The qubit layout of path graph states on each IBM Quantum devices.
    Each figure shows the qubit layout of 65-qubit devices and 27-qubit devices, respectively.
    The numbers on the figure represent the positions of virtual qubits.
  }
\end{figure}

\begin{figure}[htb]
  \begin{minipage}[b]{0.3\linewidth}
    \centering
    \includegraphics[keepaspectratio, width=\linewidth]{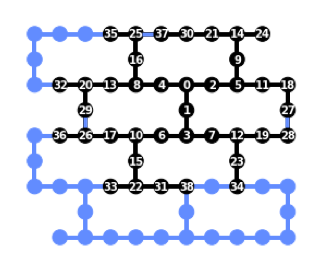}
  \end{minipage}
  \begin{minipage}[b]{0.3\linewidth}
    \centering
    \includegraphics[keepaspectratio, width=\linewidth]{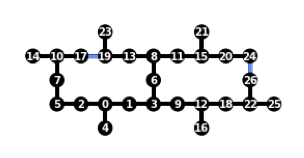}
  \end{minipage}
  \caption{ \label{fig:initial_layout_star}
    The qubit layout of path graph states on each IBM Quantum devices.
    Each figure shows the qubit layout of 65-qubit devices and 27-qubit devices, respectively.
    The numbers on the figure represent the positions of virtual qubits.
  }
\end{figure}

\begin{figure}[htb]
  \begin{minipage}[b]{0.27\linewidth}
    \centering
    \includegraphics[keepaspectratio, width=\linewidth]{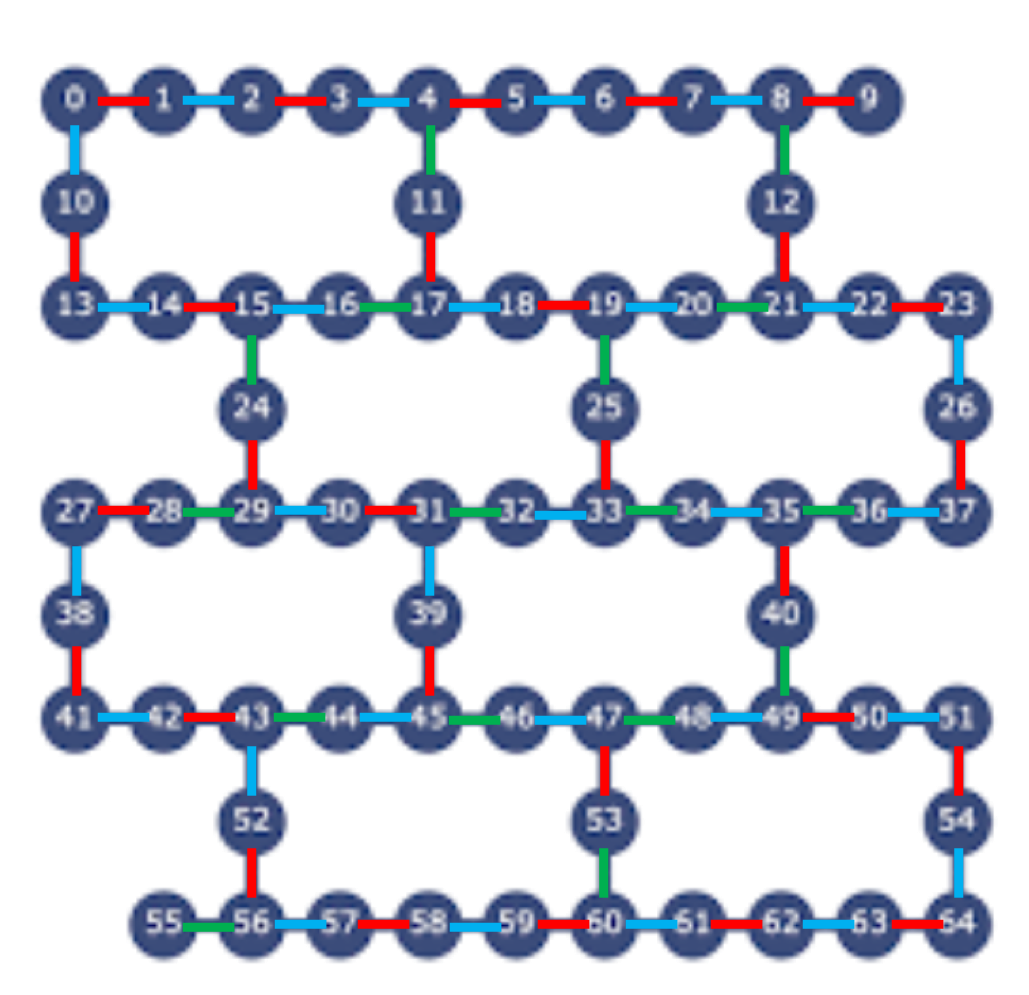}
  \end{minipage}
  \begin{minipage}[b]{0.27\linewidth}
    \centering
    \includegraphics[keepaspectratio, width=\linewidth]{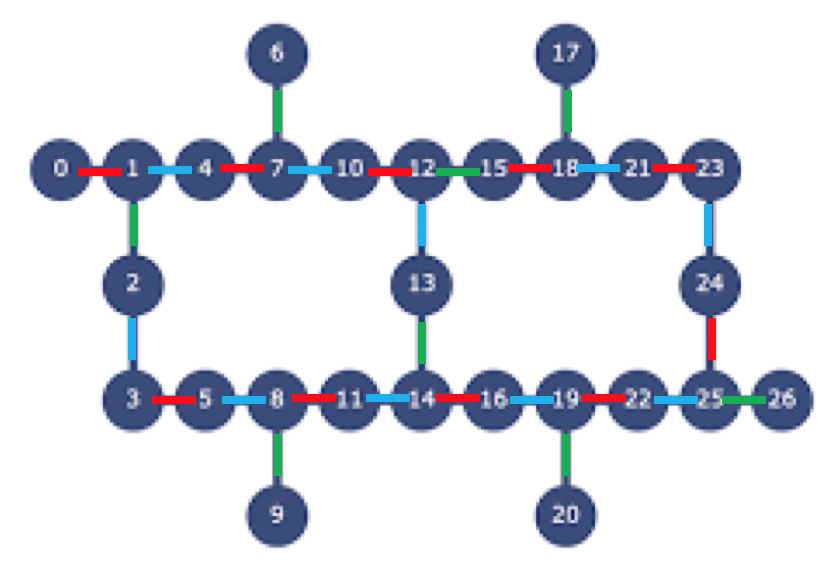}
  \end{minipage}
  \caption{ \label{fig:initial_layout_connection}
    The qubit layout of the connection graph states on each IBM Quantum device.
    Each figure shows the qubit layout of 65-qubit devices and 27-qubit devices, respectively.
    The numbers on the figure represent the positions of virtual qubits.
    In order to prepare the subdivided honeycomb graph with the whole qubits on the device, we first apply control-Z gate on the red edges, then on the blue edges, and finally on the green edges. 
    The "focused" qubits in the Baccari et al.'s inequality are set to [1,6,8,12,17,19,23,26] for 27-qubit devices, and [3,6,9,10,17,21,24,25,26,31,35,38,44,47,54,56,59,62] for the 65-qubit device.
  }
\end{figure}

\newpage

\section{Device Information of IBM Quantum Experience \label{sec:appendix_device_info}}

\begin{figure}[htb]
  \begin{minipage}[b]{0.35\linewidth}
    \centering
    \includegraphics[keepaspectratio, width=\linewidth]{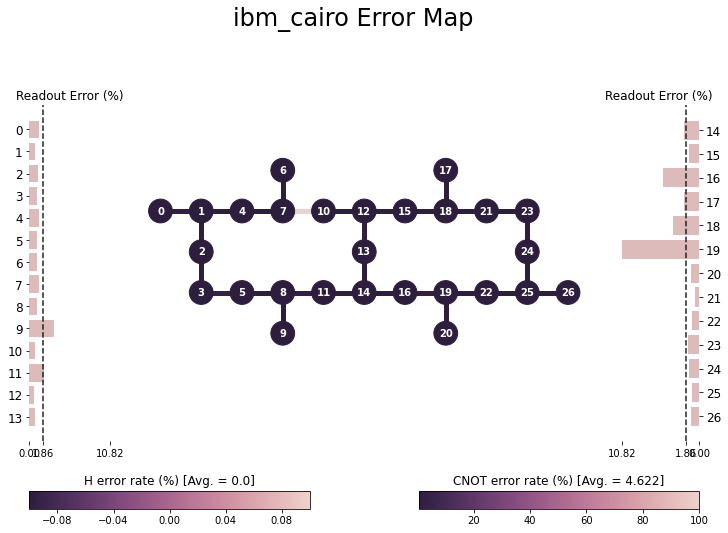}
  \end{minipage}
  \begin{minipage}[b]{0.35\linewidth}
    \centering
    \includegraphics[keepaspectratio, width=\linewidth]{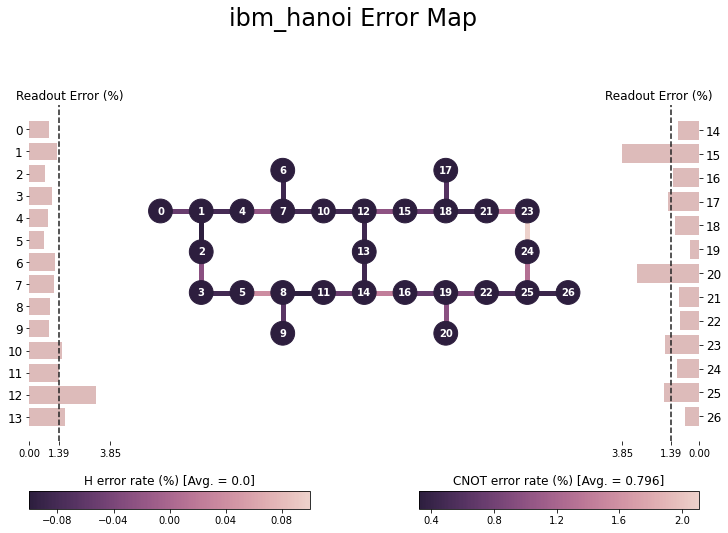}
  \end{minipage}
  \begin{minipage}[b]{0.35\linewidth}
    \centering
    \includegraphics[keepaspectratio, width=\linewidth]{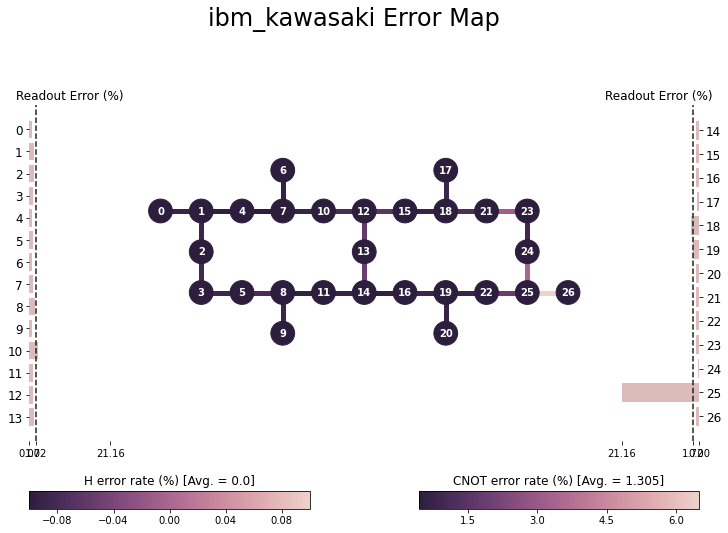}
  \end{minipage}
  \begin{minipage}[b]{0.35\linewidth}
    \centering
    \includegraphics[keepaspectratio, width=\linewidth]{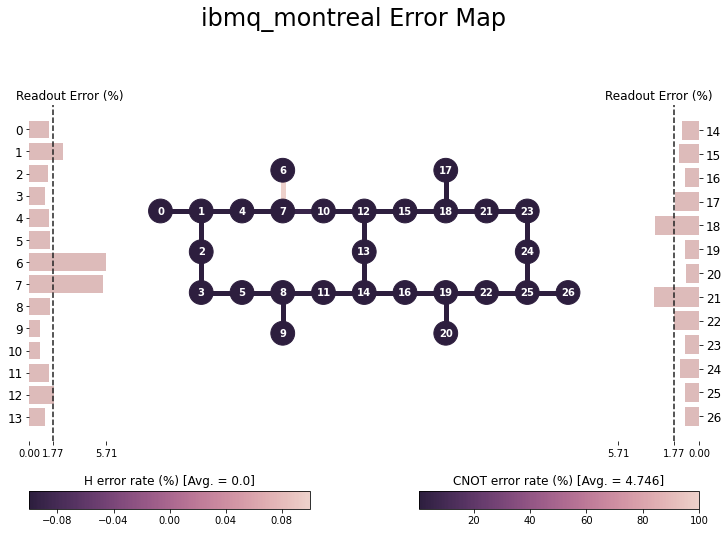}
  \end{minipage}
  \begin{minipage}[b]{0.35\linewidth}
    \centering
    \includegraphics[keepaspectratio, width=\linewidth]{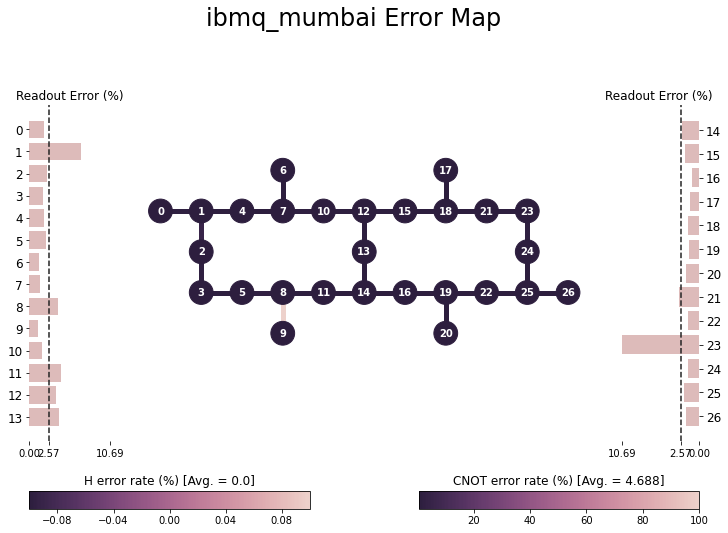}
  \end{minipage}
  \begin{minipage}[b]{0.35\linewidth}
    \centering
    \includegraphics[keepaspectratio, width=\linewidth]{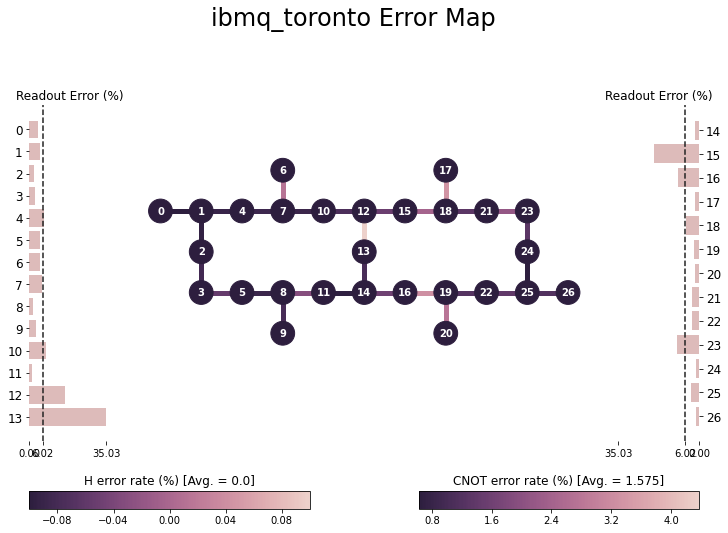}
  \end{minipage}
  \begin{minipage}[b]{0.35\linewidth}
    \centering
    \includegraphics[keepaspectratio, width=\linewidth]{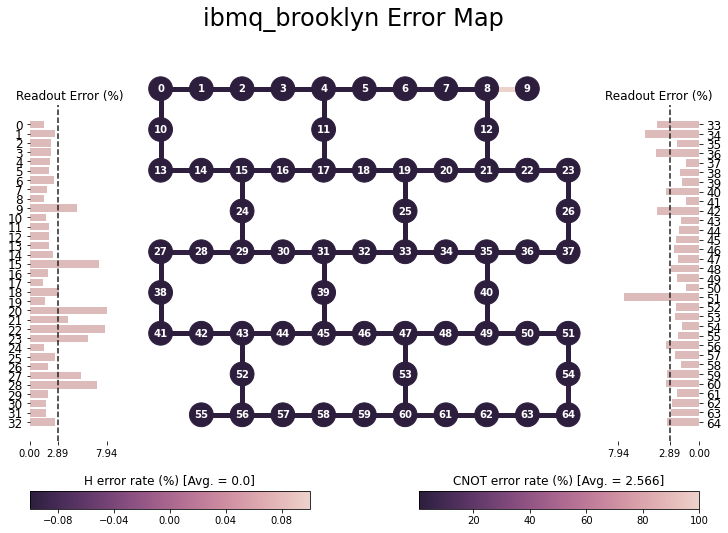}
  \end{minipage}
  \caption{ \label{fig:device_information}
    The error map of ibm\_cairo, ibm\_hanoi, ibm\_kawasaki, ibm\_montreal, ibm\_montreal, ibm\_mumbai, ibm\_toronto, and ibm\_brooklyn.
    The numbers on the figure represent the positions of physical qubits.
  }
\end{figure}

\end{document}